# A new theoretical Lorentzian Apodization function: A Simulations based Comparison with traditional Apodization functions

*Dalton H Bermudez*[1]

1 Department of Medical Physics, University of Wisconsin-Madison, Madison, WI, USA

*Correspondance:
Dalton H Bermudez
dbermudez@wisc.edu

**Keywords: Apodization, Ultrasound Imaging, Instrumentation, Signal processing, Ultrasound pressure wave**

**Abstract**

Background

Standard apodization methods suppress the side lobes at the expense of increasing FWHM, leading to a decrease in lateral resolution. Grating lobes tend to interfere with the main lobe, resulting in major artifacts in ultrasound. A simple mathematical apodization function is needed to reduce the side lobes while decreasing FWHM, resulting in an increase in the beam's lateral resolution.

Purpose

The advantage of the new apodization function is that it works as well as a Hanning function at suppressing first grating lobes while reducing the FWHM, leading to an increase in the lateral resolution of the beam. The method would potentially improve the lateral resolution while reducing ultrasound artifacts caused by interferences of the main beam with the side lobes.

Methods

Simulations of the ultrasound beam with new quadratic apodization and no apodization were modeled with a governing equation for a wave's pressure. The new apodization method was modeled with a rectangular function convoluted with a Fourier transform of an Lorenzian. The FWHM and the amplitude of the first side lobes were quantified and compared for different apodization functions.

Results

The proposed Lorentzian apodization outperformed the no apodization in FWHM, with the values being 0.906 and 1.22 mm, respectively. When comparing the suppression of the first grating lobe of the new apodization with a standard Hanning apodization, the corresponding values were -31.47 dB and -31 dB, respectively. While the corresponding B-mode image sharpness were 7.3398, 6.6009, and 7.0039 for no apodization, Hanning apodization, and Lorentzian apodization, respectively. Analysis of the ultrasound phantom images demonstrated that apodization functions exerted distinct effects on sharpness and contrast-to-noise ratio (CNR). The rectangular window produced the greatest sharpness (0.014), consistent with strong edge definition but elevated noise, whereas smoother tapers such as Hanning, Hamming, and Gaussian yielded lower sharpness values (~0.0068–0.0069). In contrast, CNR analysis of the low-contrast target showed that Lorentzian-based apodizations, particularly at parameters $a = 8$ and $a = 15$, achieved the highest CNR (>3.2), surpassing both the original reconstruction and standard windowing functions. Incorporating Slepian tapering with Lorentzian windows generally reduced CNR, suggesting that excessive broadening of main beam diminishes lesion conspicuity. Overall, these findings highlight a trade-off between sharpness and CNR, with Lorentzian apodization providing the most favorable balance for enhancing low-contrast detectability without markedly compromising resolution.

Conclusion

The new proposed apodization improves the FWHM compared to a rectangular apodization and outperforms the Hanning apodization functions in suppressing the amplitude of the first side lobe. The new approach with the Slepian sequences also gives superior CNR compared to standard apodizations at a slight expense to the image sharpness.

## Introduction

Ultrasound imaging has emerged as a non-invasive and widely used medical imaging modality, allowing clinicians to visualize internal organs and tissues in real-time. The diagnostic utility of ultrasound largely depends on the quality and resolution of the acquired images. Over the years, advancements have been made to enhance ultrasound imaging, and one such technique is apodization. Apodization is a signal processing technique employed to improve the quality of ultrasound images by mitigating undesirable artifacts and enhancing spatial resolution. This article aims to introduce a new theoretical apodization-based function which unlike most apodization-based functions dramatically reduces the side lobes from the variation in pressure profile in ultrasound imaging while enhancing the lateral resolution by reducing the full width at half maximum of the main lobe.

Ultrasound imaging has revolutionized the field of medical diagnostics, providing non-invasive and real-time visualization of internal structures and organs [1]. Since its inception in the early 1950s, ultrasound has become an essential tool for clinicians in various medical specialties, including obstetrics, cardiology, radiology, and many others [1-2]. The widespread use of ultrasound can be attributed to its safety, affordability, portability, and ability to generate high-resolution images [2].

Despite the numerous advantages, conventional ultrasound imaging techniques face certain limitations that can compromise image quality and diagnostic accuracy. One such limitation is reduced CNR , which can obscure important anatomical details and lead to misinterpretation of findings [3].

In recent years, progress has been made in the field of ultrasound signal processing, aiming to mitigate artifacts and enhance image quality. One prominent technique that has gained attention is apodization [4]. Apodization involves the application of mathematical windowing functions to the ultrasound signals before the beamforming process [4]. This approach selectively reduces the amplitude of the signals at the edges, which helps in mitigating artifacts such as sidelobes [4]. By doing so, apodization enhances spatial resolution and improves image sharpness, leading to more accurate diagnoses and better patient outcomes [5].

The use of apodization has become increasingly prevalent in modern ultrasound imaging systems, allowing clinicians to obtain clearer and more precise images of the structures of interest. Various apodization algorithms have been developed, each tailored to address specific imaging challenges and clinical requirements [6]. Researchers continue to explore innovative apodization techniques to strike a balance between improving image quality and preserving the signal-to-noise ratio.

In conclusion, apodization stands as a promising signal processing technique in ultrasound imaging, offering potential to improve image quality and resolution. The integration of apodization into current ultrasound systems has shown promising results in improving Contrast to Noise ratio (CNR) and enhancing diagnostic accuracy. With continuous research and innovation, apodization is expected to remain a key area of interest for researchers and clinicians alike, driving the advancement of ultrasound technology and ultimately benefiting patients worldwide. This paper explores the implementation of a new theoretical apodization-based function for the variation of pressure of ultrasound waves.

**Theory/Methods**

We will first derive the pressure profile of the ultrasound with a no apodization function. We start with the following governing equation for variation of pressure in ultrasound (1) [9]. Where P(x,z,$\omega$) is the acoustic pressure field at spatial point (x,z) for a specific angular frequency $\omega$, $\rho_0$ is the ambient density of the medium, $V_0(\omega)$ is the frequency dependent source velocity, Z is the lateral coordinate, x is the lateral coordinate, $\lambda$ is the wavelength of the ultrasound wave, $e^{-i\frac{\pi}{\lambda z}[z^2+x^2]}$ is the phase shift term where $x^2+z^2$ represents the distance to the transducer to a point (x,z) in the medium and -i introduces the phase modulation due to the wave propagation, and $A(\varsigma_x)$ is the aperture function.

$$P(x,z,\omega) = \sqrt{\frac{i\omega\rho_0 V_0(\omega)}{2\pi z}}\lambda z e^{-i\frac{\pi}{\lambda z}[z^2+x^2]} \int A(\varsigma_x) e^{i2\pi(x\varsigma_x)} d\varsigma_x \tag{1}$$

The no apodization function we are going to insert at A($\mathcal{J}_x$) is (2). Where $\Pi$ is a rectangular function with width L,  L is the length of the active aperture.

$$A_{rect}(x_0) = \Pi\left(\frac{x_0}{L}\right) \tag{2}$$

Applying equation (2) to equation (1) results in the following (3).

$$P(x,z,\omega) = \sqrt{\frac{i\omega\rho 0V0(\omega)}{2\pi z}}\lambda z e^{-\frac{i\pi}{\lambda z}[z^2+x^2]} \int \Pi\left(\frac{\zeta x}{L}\right) e^{i2\pi(x\zeta x)} d\zeta x \tag{3}$$

The inverse Fourier transform of a rectangular function is a sinc which results in an equation (4)

$$\sqrt{\frac{i\omega\rho 0V0(\omega)}{2\pi z}}\lambda z e^{-\frac{i\pi}{\lambda z}[z^2+x^2]} \text{Lsinc}\left(\frac{xL}{2}\right) \tag{4}$$

Now normalizing with P(0,z,w) results in the following (5)

$$\frac{P(x,z,\omega)}{P(0,z,w)} = \frac{\left(\sqrt{\frac{i\omega\rho 0V0(\omega)}{2\pi z}}\lambda z e^{-\frac{i\pi}{\lambda z}[z^2+x^2]} L\sin c\left(\frac{xL}{2}\right)\right)}{\sqrt{\frac{i\omega\rho 0V0(\omega)}{2\pi z}}\lambda z e^{-\frac{i\pi}{\lambda z}[z^2+(0)^2]} L\sin c(0)} = e^{-\frac{i\pi x^2}{\lambda z}} \sin c\left(\frac{xL}{2}\right) \tag{5}$$

The new theoretical apodization based function we will introduce is (6)

$$A(x_0) = \Pi\left(\frac{x_0}{L}\right) ** FT\{\frac{1}{2x^2+1}\} \quad (6)$$

Were equation (6) is a convolution between a rectangular function and the Fourier transform of $\frac{1}{2x^2+1}$. Applying equation (1) to equation (6) results in the following (7)

$$P(x,z,w) = \sqrt{\frac{iwp0Vo(w)}{2\pi z}}\,\lambda z e^{-\frac{i\pi}{\lambda z}[z^2+x^2]} \int \Pi(\frac{\zeta x}{L}) ** [\int \frac{1}{2x^2+1}\, e^{-i2\pi(x\zeta x)}\, dx]\, e^{i2\pi(x\zeta x)} d\zeta x \quad (7)$$

The inverse Fourier transform of a convolution operation is the product between the two functions resulting in (8). The inverse Fourier transforms of a rectangular function is a sinc and the inverse Fourier transforms of the Fourier transforms of an Lorentzian is just the Lorentzian function.

$$\sqrt{\frac{iwp0Vo(w)}{2\pi z}}\,\lambda z e^{-\frac{i\pi}{\lambda z}[z^2+x^2]}\, sinc(\frac{xL}{2})\frac{1}{2x^2+1} \quad (8)$$

Now normalizing with P(0,z,w) results in the following (9)

$$e^{-\frac{i\pi x^2}{\lambda z}} sinc\left(\frac{xL}{2}\right)\frac{1}{2x^2+1} \quad (9)$$

Simulations of ultrasound imaging using a wire phantom were conducted utilizing the Field II MATLAB toolbox. A linear transducer array with 128 elements was modeled, operating at a center frequency of 9 MHz and a wavelength of 0.17 mm. In the simulations the extent of the depth was of 40 mm. The element pitch, width, and kerf were set at fractions of the wavelength to ensure proper element spacing and minimal grating lobes. The transducer apertures were constructed for both transmit and receive operations, with focusing delays manually implemented for precise control over beamforming.

A four-cycle sinusoidal pulse modulated by a Hanning window was transmitted, with scatterers strategically positioned at defined axial and lateral coordinates to simulate the wire phantom. Field II's calc_scat_multi function was used to generate raw radiofrequency (RF) data from the scatterers, which were then processed using delay-and-sum beamforming.

To enhance lateral resolution and suppress sidelobes, various apodization functions, including uniform, Hanning, and Lorentzian weighting, were applied during the receive beamforming process. The Lorentzian weighting can be represented by equation 10 (Figure 1), where N_elements in the total number of piezoelectric elements in the model transducer, x is an array from 0 to the total number of elements in the model array, and a is a constant term which sets the spread of the Lorentzian which can be adjusted to adjust the level of suppression of side lobes. In the Field II simulation, a was set to 30 for the wire phantom simulation but a value of 15 was used for the contrast phantom simulations. For each case, the time-of-flight delays for each element were calculated based on the distances between the active transmit and receive elements and the scatterers. The received RF signals were interpolated using spline interpolation, and their contributions were summed to create the final beamformed image.

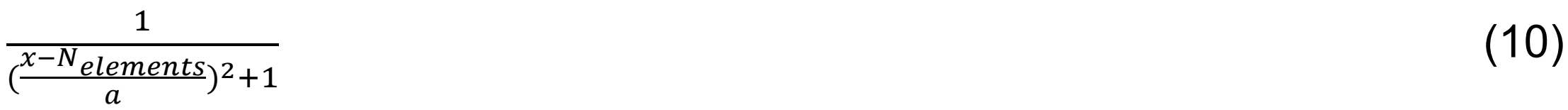

$$\frac{1}{(\frac{x-N_{elements}}{a})^2+1} \qquad (10)$$

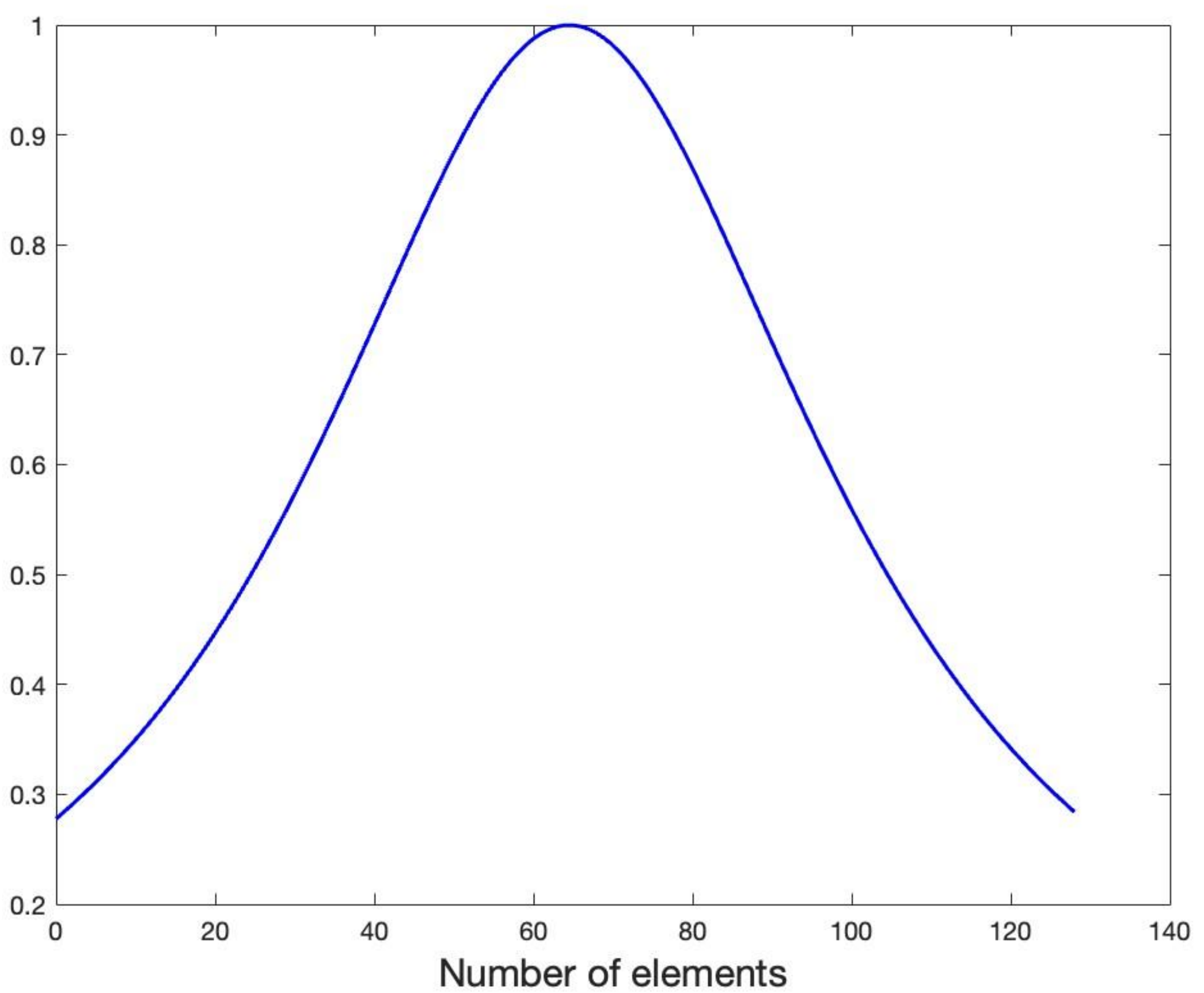

**Figure 1:** Illustration of proposed Lorentzian apodization function used for Field II simulations.

The envelope of the beamformed image was extracted using the Hilbert transform, followed by log compression (20×log10 of the envelope) to generate the B-mode image. The resulting images were normalized to ensure that the peak signal corresponded to 0 dB. Visualization included overlays of the transmit and receive element positions and scatterers for clarity. This methodology allowed for a detailed evaluation of the effects of different apodization schemes on image quality and resolution in the simulated ultrasound environment. For quantifying the sharpness of the simulated B-mode scan with wire phantom a gradient based method was used. To calculate the CNR rectangular circular ROI regions were drawn one in low contrast region and the other one in the background (low intensity region) of the B-mode image. The CNR was calculated based on the mean intensities of the background and signal ROI region and the standard deviation of the background ROI.

Contrast phantom was imaged with B-mode with a 10L4 transducer, then the original image was subjected to different apodization. Using Field II, the point spread function for an 10L4 transducer was simulated for all the different apodization functions to then apply in Fourier space with the original image. The Contrast to Noise ratios of the contrast phantoms were obtained as well as the mean signal of a low contrast region in the phantom compared to surrounding background signal. To prevent artificial CNR inflation due to side lobes a smaller and more isolated ROIs were selected.

**Results**

The extra component of the new apodization function is as follows (**Figure 2)**:

$$\frac{1}{2x^2+1}$$

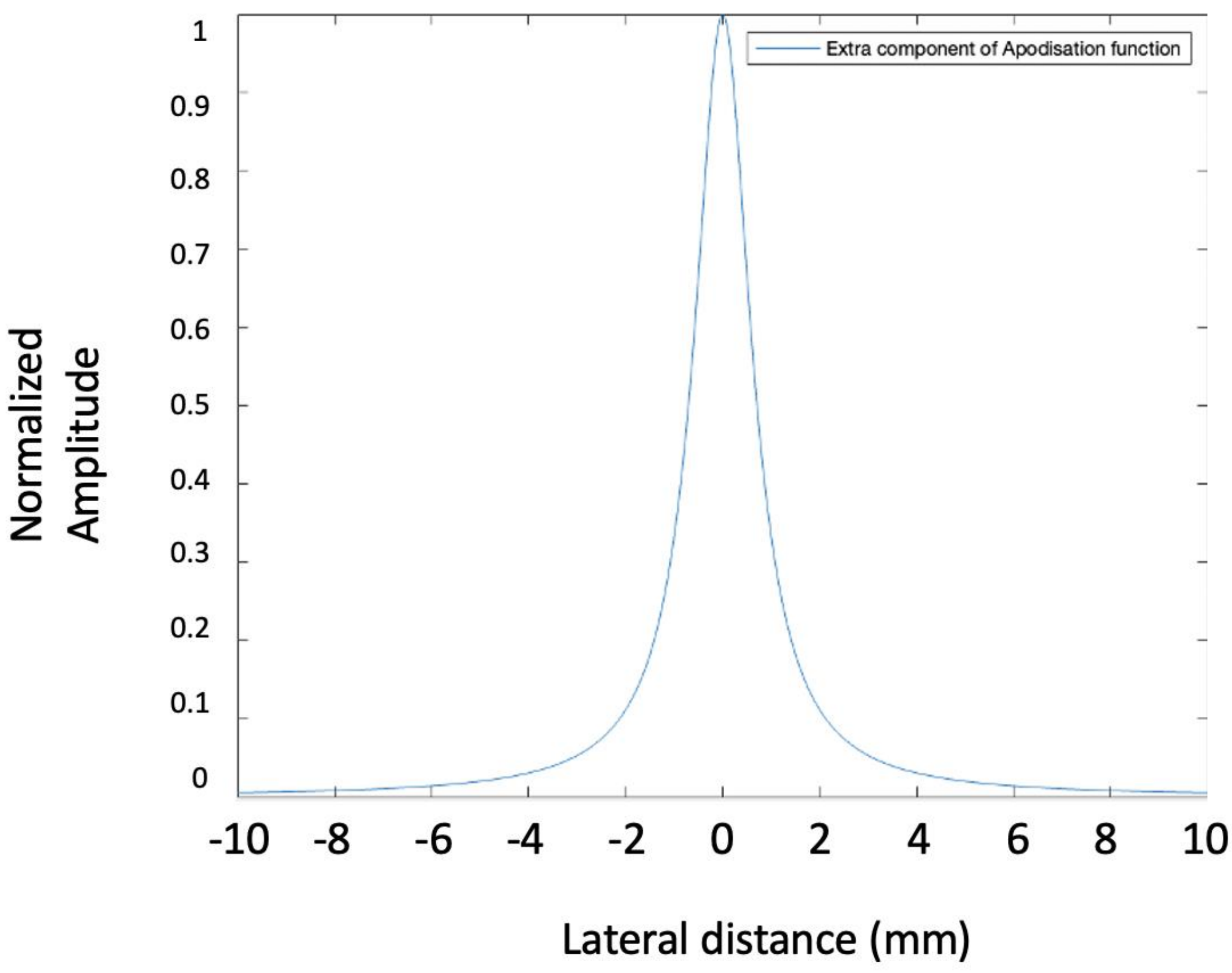

**Figure 2**
A plot of the part of the apodization function which regulates the side lobes and increases the lateral resolution of the ultrasound beam.

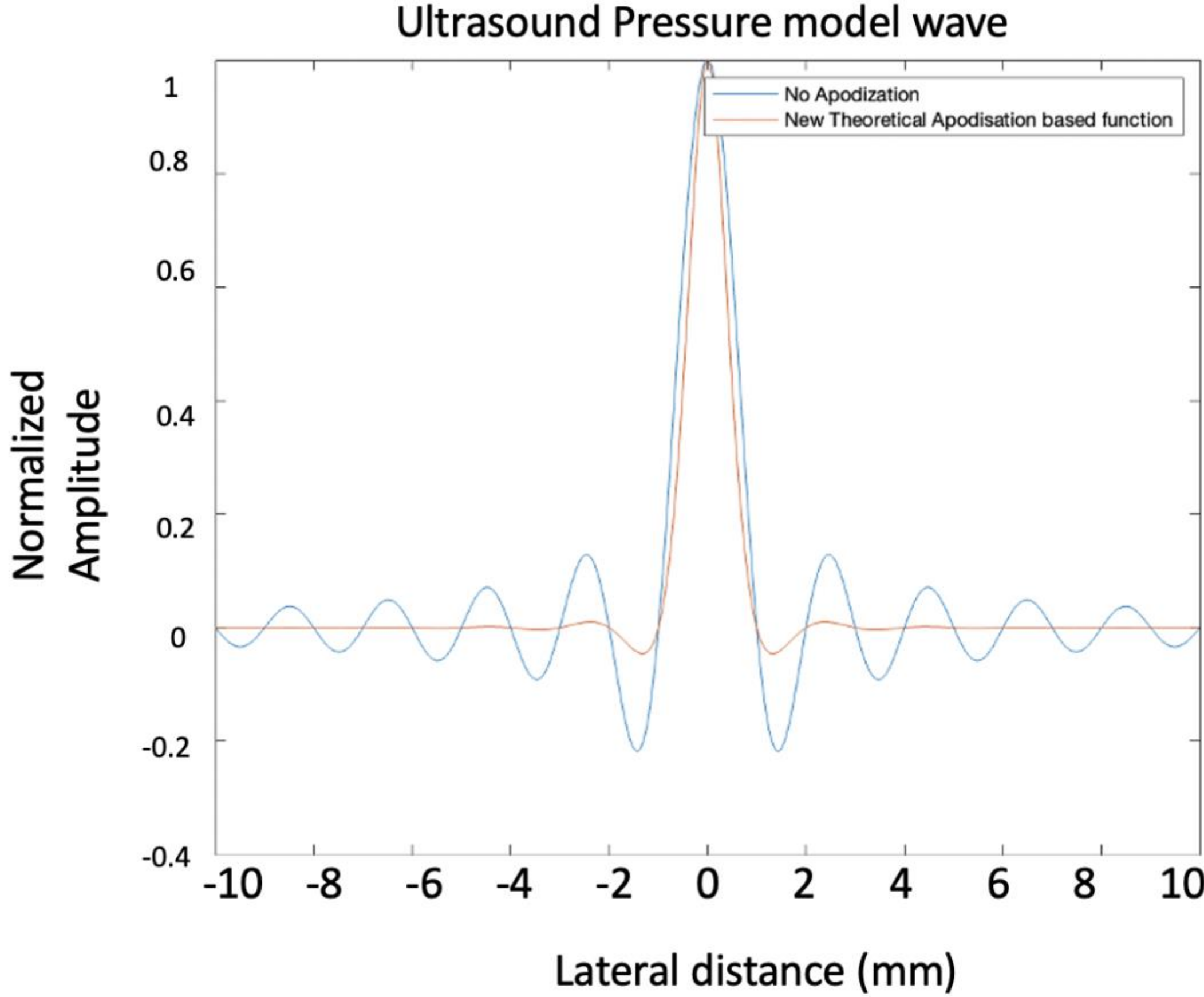

**Figure 3**

The blue plot shows the variation in pressure of an ultrasound wave with no apodization. While the red plot shows the variation in pressure of the ultrasound wave with the new theoretical-based apodization function.

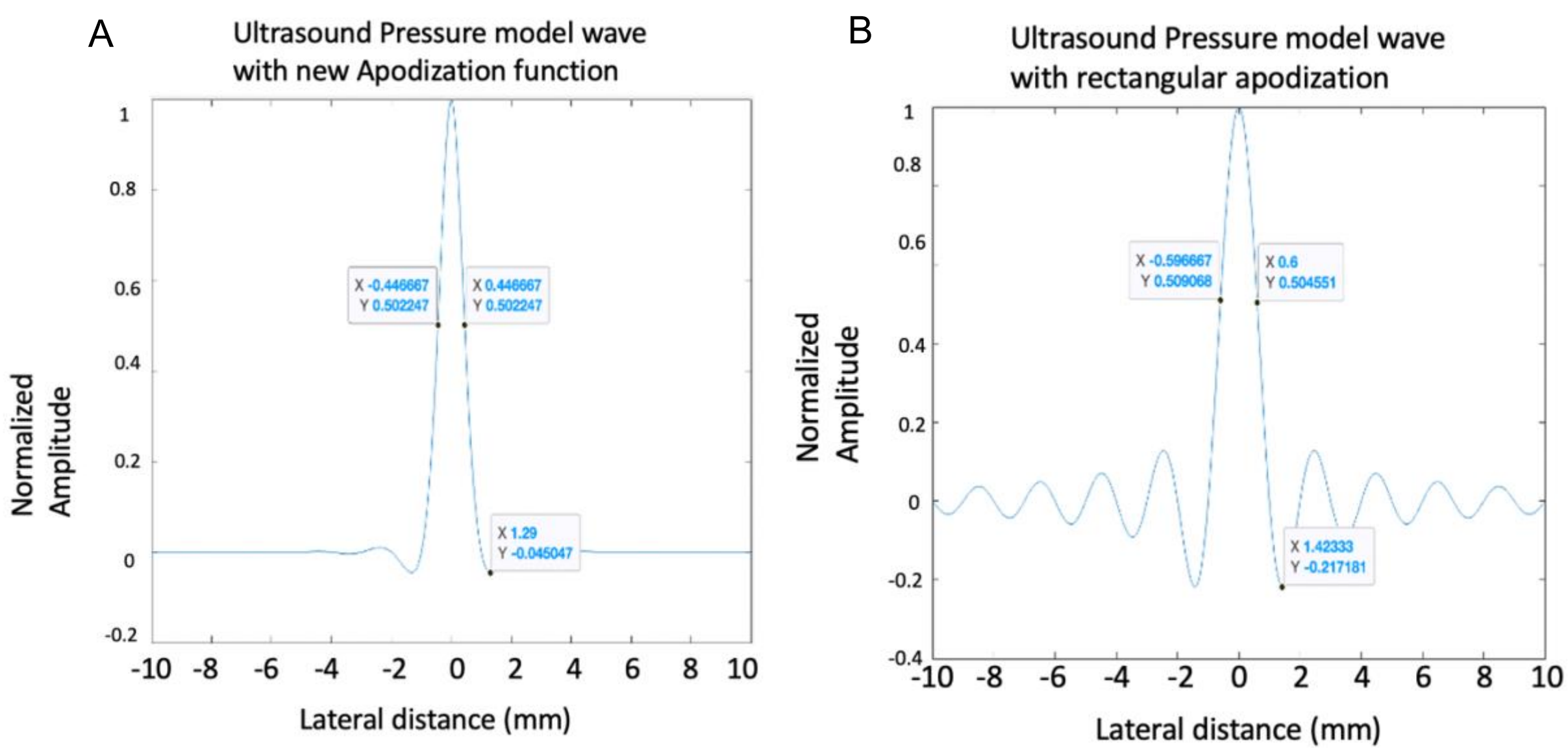

**Figure 4**
**(A)** Plot the variational wave profile for no apodization and **(B)** new theoretical apodization with points at the full-width half maximum and amplitude values for the first side lobe.

The Full-width half maximum for the no apodization variational pressure profile is 1.22 mm, while the Full-width half maximum for the new apodization function is 0.906 mm **(Figure 3)**. The amplitude for the 1st grating lobe for the no apodization is -15.61 db, while the amplitude for the 1st grating lobe for the new apodization-based function is -31.4656 db **(Figure 4)**. This shows that this new inverted quadratic apodization function performs better at suppressing the first side lobe than the Hanning apodization function while not widening the main lobe of the ultrasound beam **(Table 1b)**. Although the mathematics of apodization in ultrasound is well known, the introduction of an inverted quadratic function allows the proper reduction of side lobes in the ultrasound beam while reducing the full-width half maximum of the main lobe. This results in an enhancement of the lateral resolution of ultrasound. While most currently introduced apodization functions result in an increment of the full-width half maximum of the main lobe in exchange for reducing the grating lobes of the ultrasound beam **(Table 1a)**.

| | **A)** | **B)** |
|---|---|---|
| Apodization function | Main Beam FWHM (mm) | Suppression of first side lobe |
| Uniform (Rectangular) | 1.21 | -13dB |
| Hanning | 2.00 | -31dB |
| Hamming | 1.81 | -40dB |

**Table 1**

(a) Effects of various apodization functions on FWHM (b) Effects of the various apodization functions on the suppression of the first side lobes.

***Simulation with Field II on wire phantom***
***Point Spread Function Simulations:***

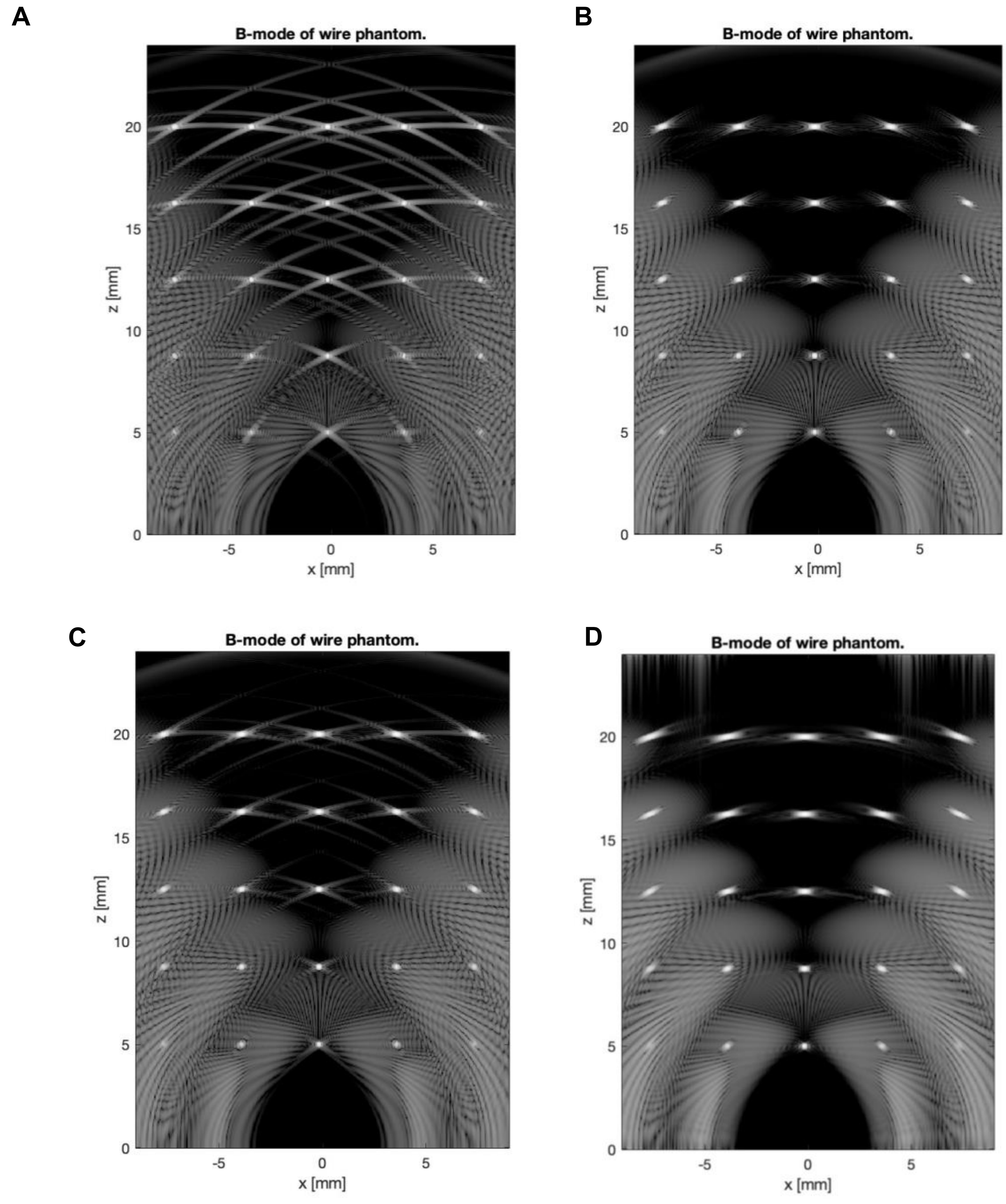

**Figure 5:** Simulations of wire phantom B-mode image with the following apodizations: A) No apodization (rectangular function), B) Hanning Apodization functions, C) proposed

Lorentzian function, and D) Lorentzian function with Slepian sequences to reduce side lobe energy

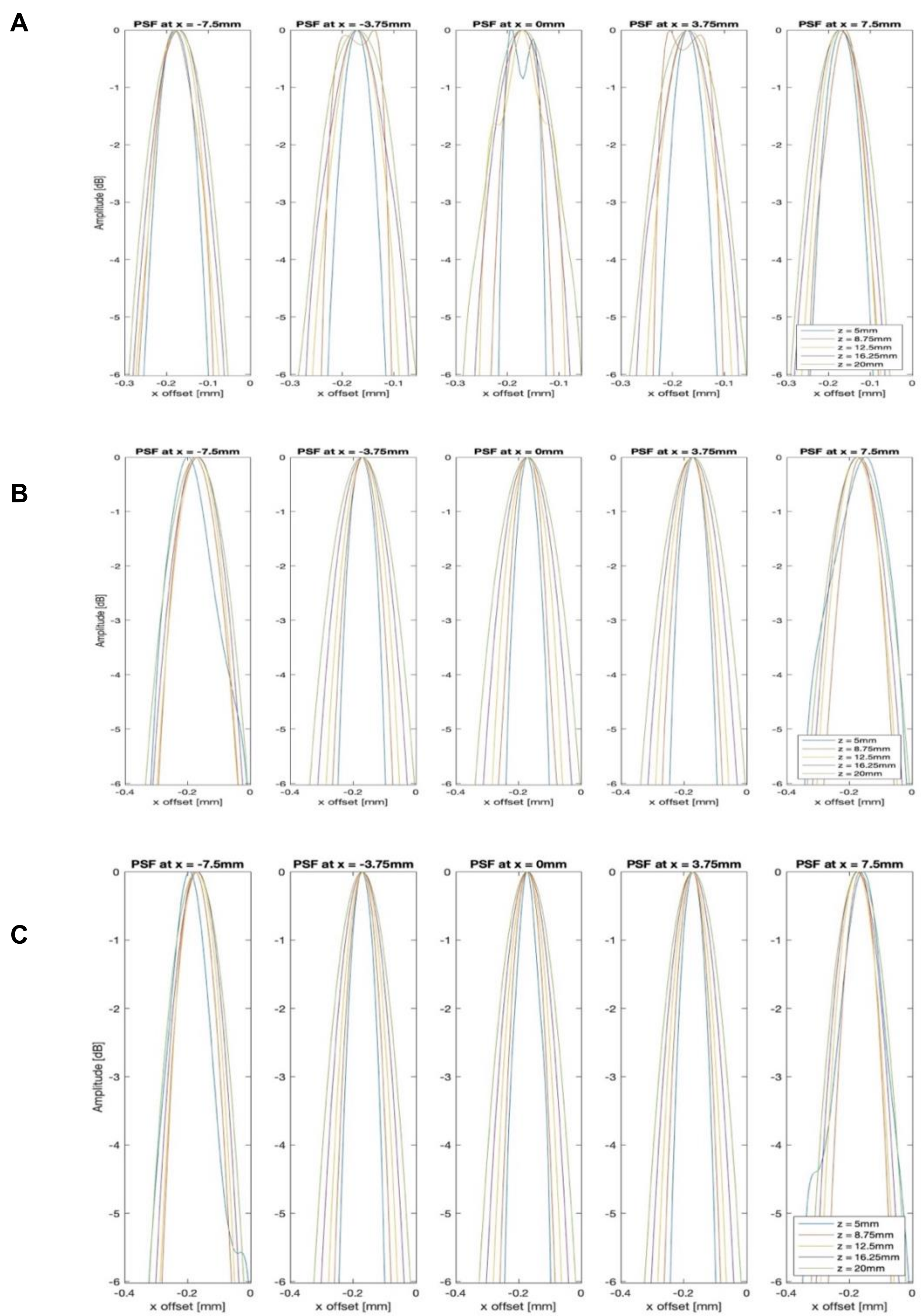

**Figure 6: A) PSF no apodization B) PSF with Hanning apodization C) PSF with Proposed Lorentzian apodization**

The images represent simulated B-mode ultrasound scans of a wire phantom (Figure 5), each generated using different apodization functions. Apodization, a key beamforming technique, is applied to reduce sidelobe artifacts and enhance the quality of ultrasound images. The wire phantom consists of discrete scatterers designed to evaluate the imaging system's resolution and artifact suppression capabilities.

In the image without apodization, all elements of the ultrasound transducer are weighted equally. This uniform weighting produces high-intensity main lobes but also results in reduced sidelobe artifacts. These sidelobes manifest as repetitive interference patterns around the scatterers, reducing the contrast and making it harder to distinguish discrete scatterers clearly. The lack of apodization results in a wider main lobe, leading to reduced lateral resolution and higher noise artifacts. The corresponding Sharpness for the simulated B-mode image with no apodization of the wire phantom was of 7.3398.

The Hanning apodization applies a smooth, tapered weighting across the transducer elements, with values decreasing gradually from the center to the edges. This tapering suppresses sidelobes, resulting in improved image contrast and reduced noise around the scatterers. The lateral resolution is enhanced as the beam profile becomes narrower. However, the main lobe intensity is slightly reduced compared to the no-apodization case, which is an expected tradeoff when sidelobes are suppressed. The corresponding Sharpness for the simulated B-mode image with Hanning apodization of the wire phantom was of 6.6009.

Proposed Lorentzian apodization applies a steeper weighting, giving higher emphasis to central elements and progressively reducing the contribution of outer elements. This results in a highly focused beam with a narrower main lobe and minimal sidelobe artifacts. The scatterers are resolved with excellent contrast, and the suppression of sidelobes is more pronounced than in the Hanning apodization case. However, the sharper focus may lead to reduced sensitivity to signals at the periphery of the imaging field The corresponding Sharpness for the simulated B-mode image with the Lorentzian of the wire phantom was of 7.0039.

*Contrast from B-mode acquired phantom image :*

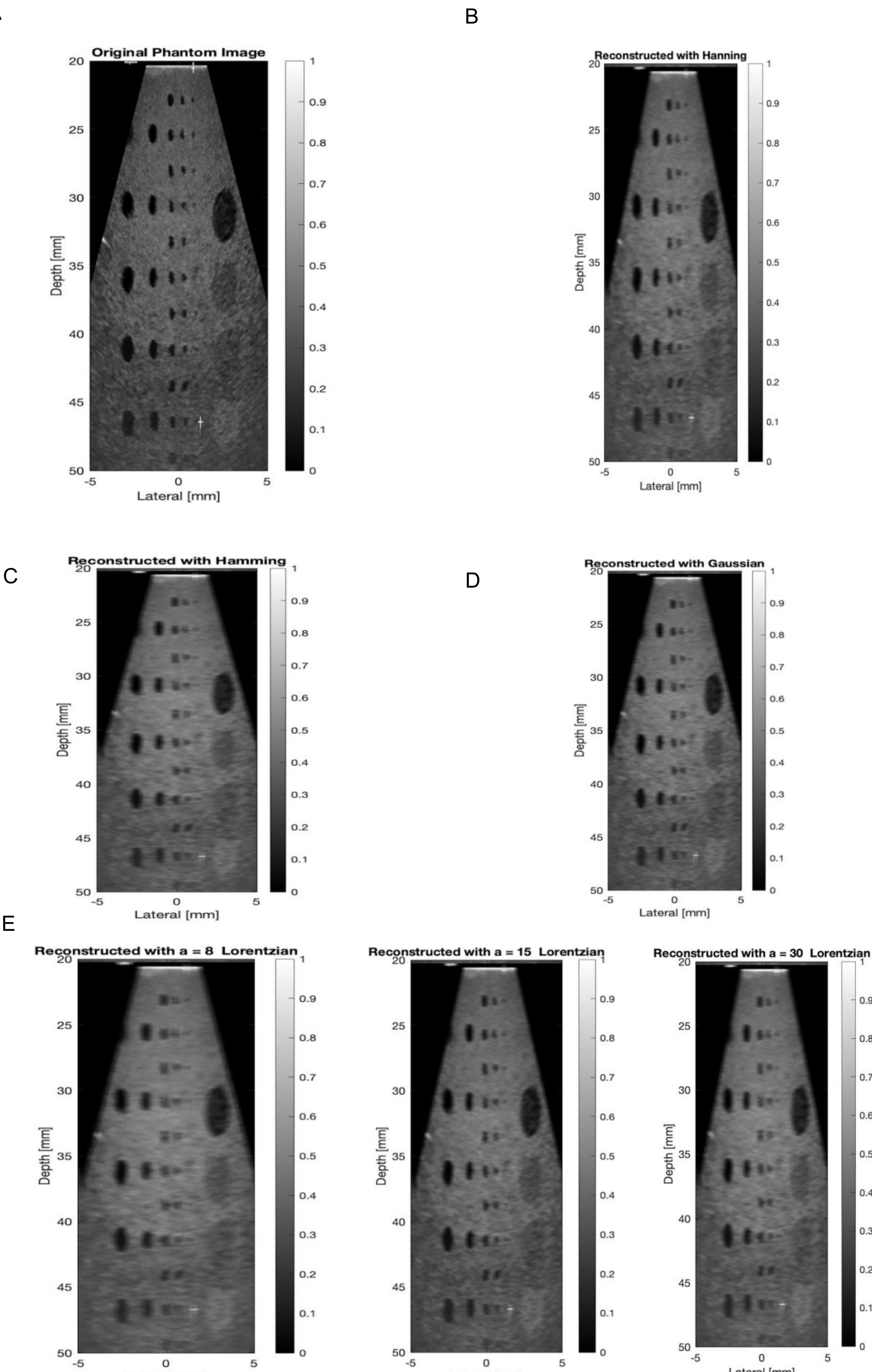

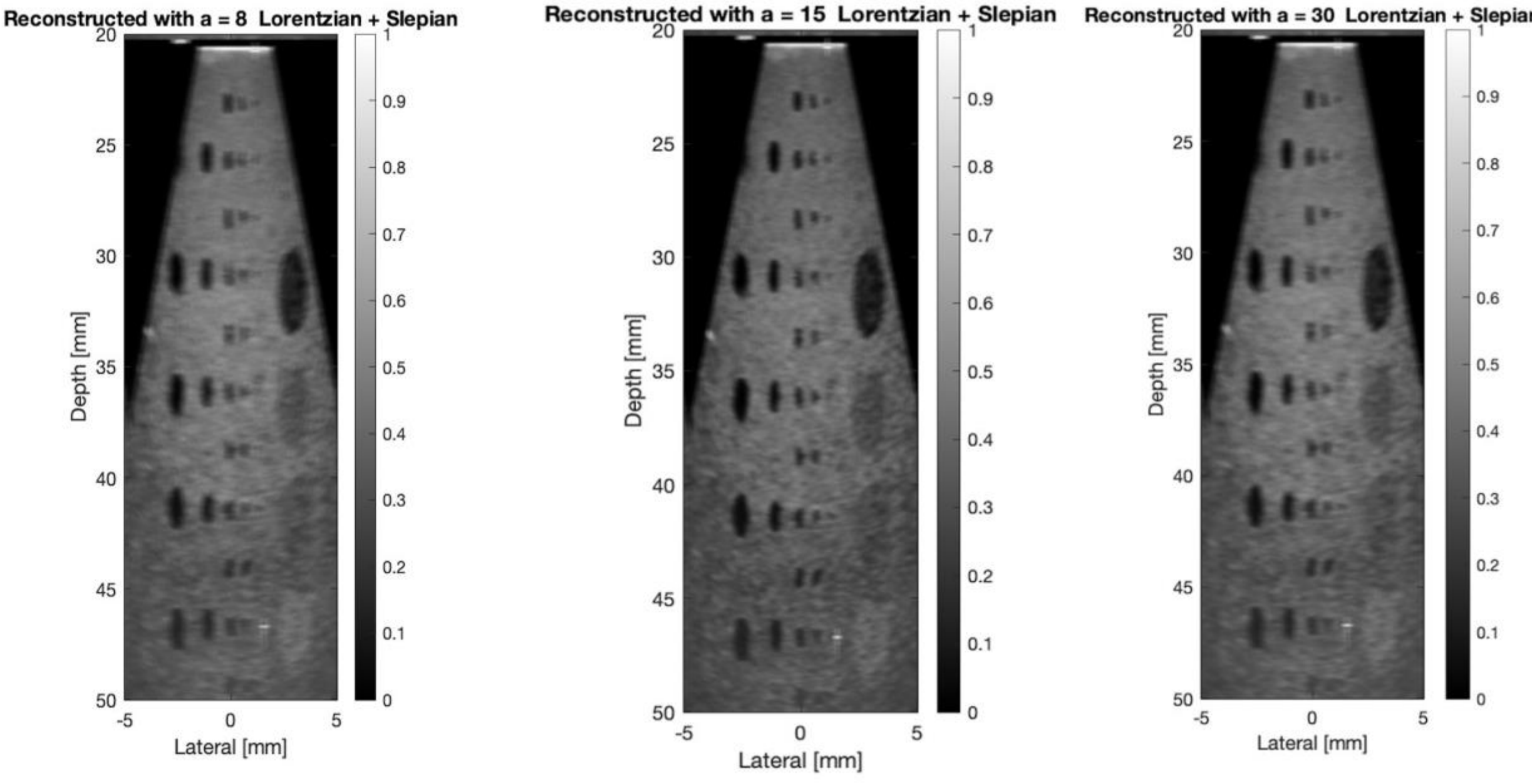

**Figure 7:** contrast phantom using A) Rectangular (original acquired image), B) Hanning, C) Hamming, D) Gaussian and E) Lorentzian at apodization various beam width (a = 8, 15, 30) apodization functions. F) Lorentzian + Slepian at apodization various beam width (a = 8, 15, 30)

A

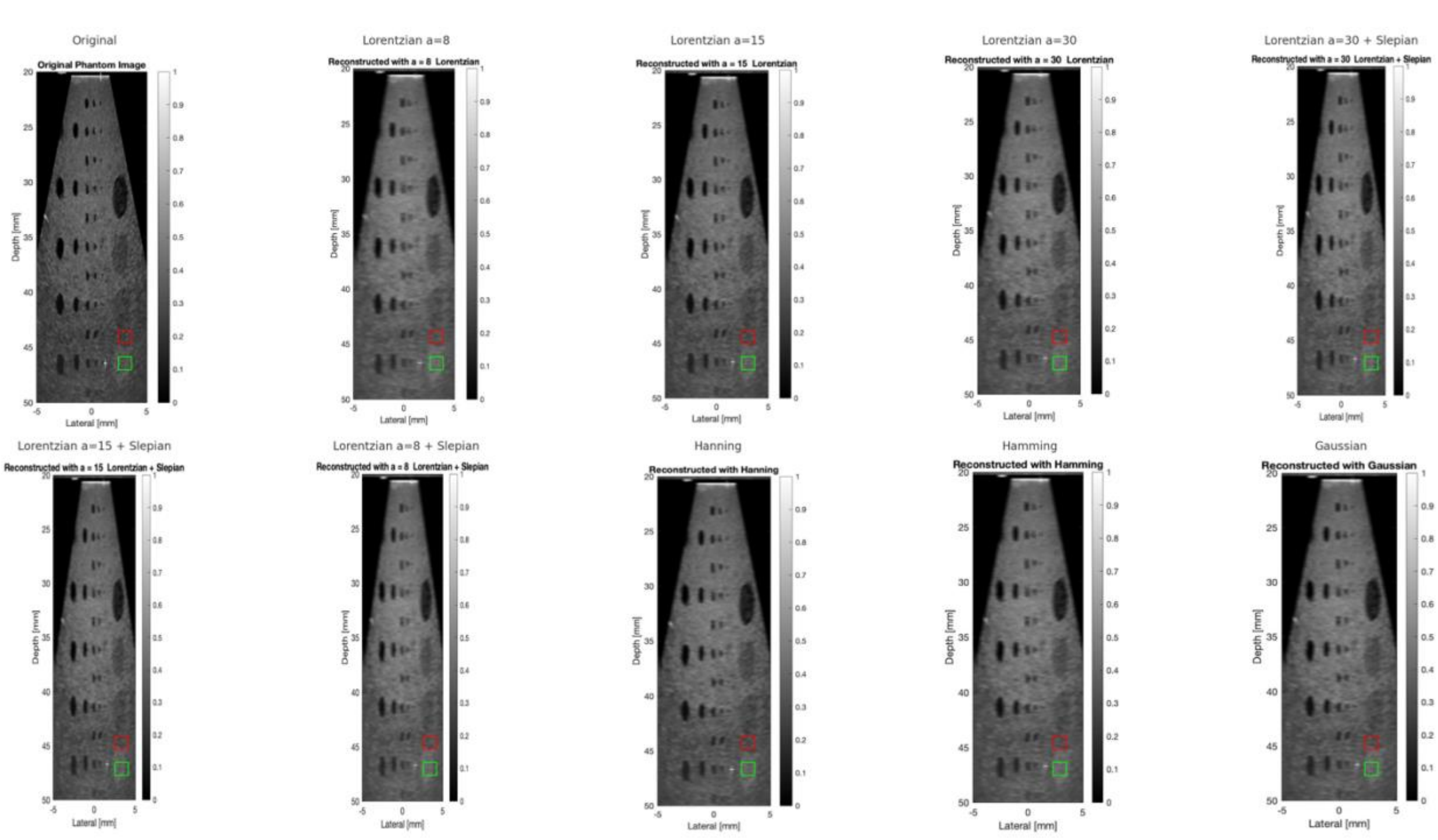

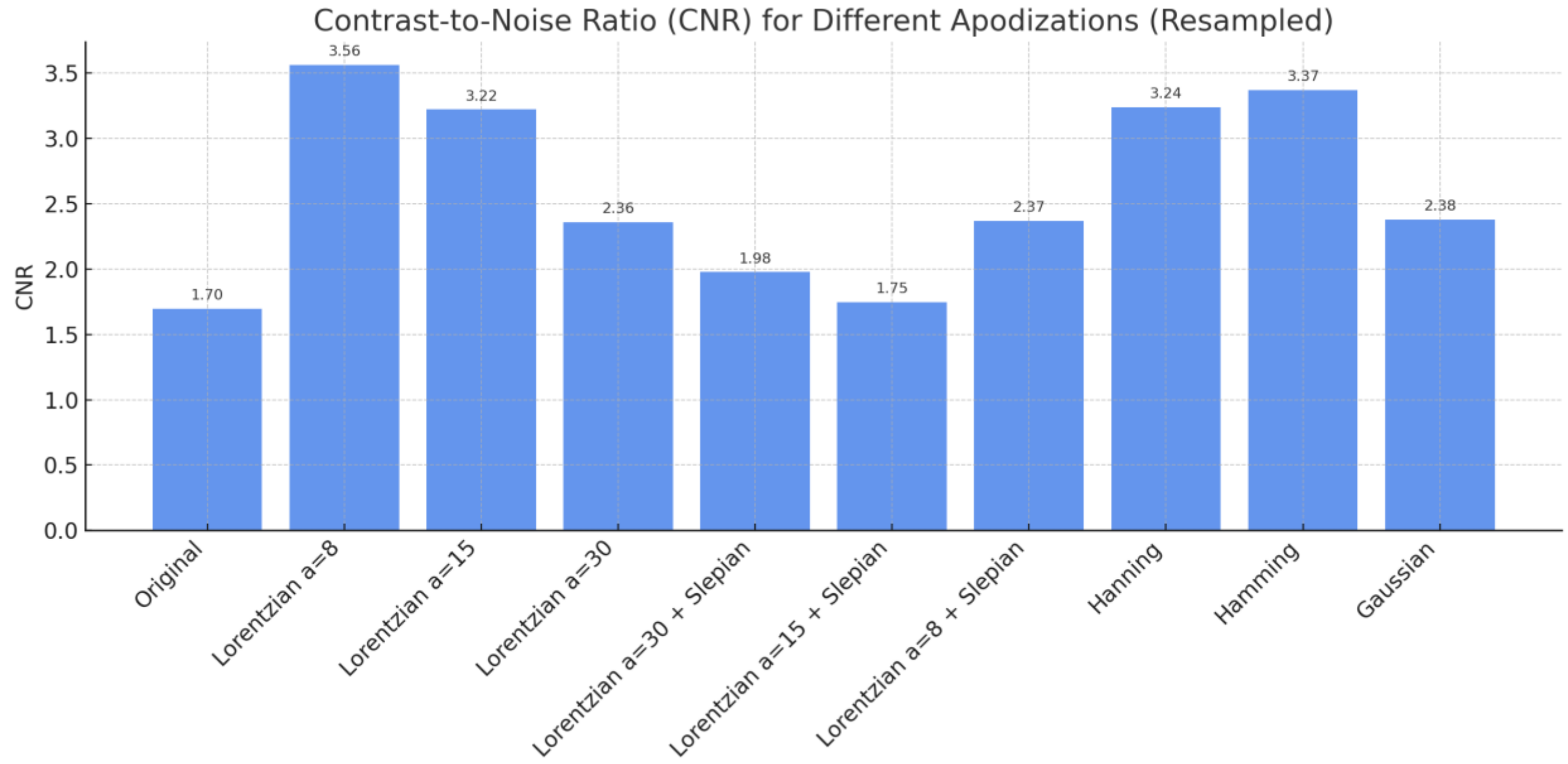

**B**

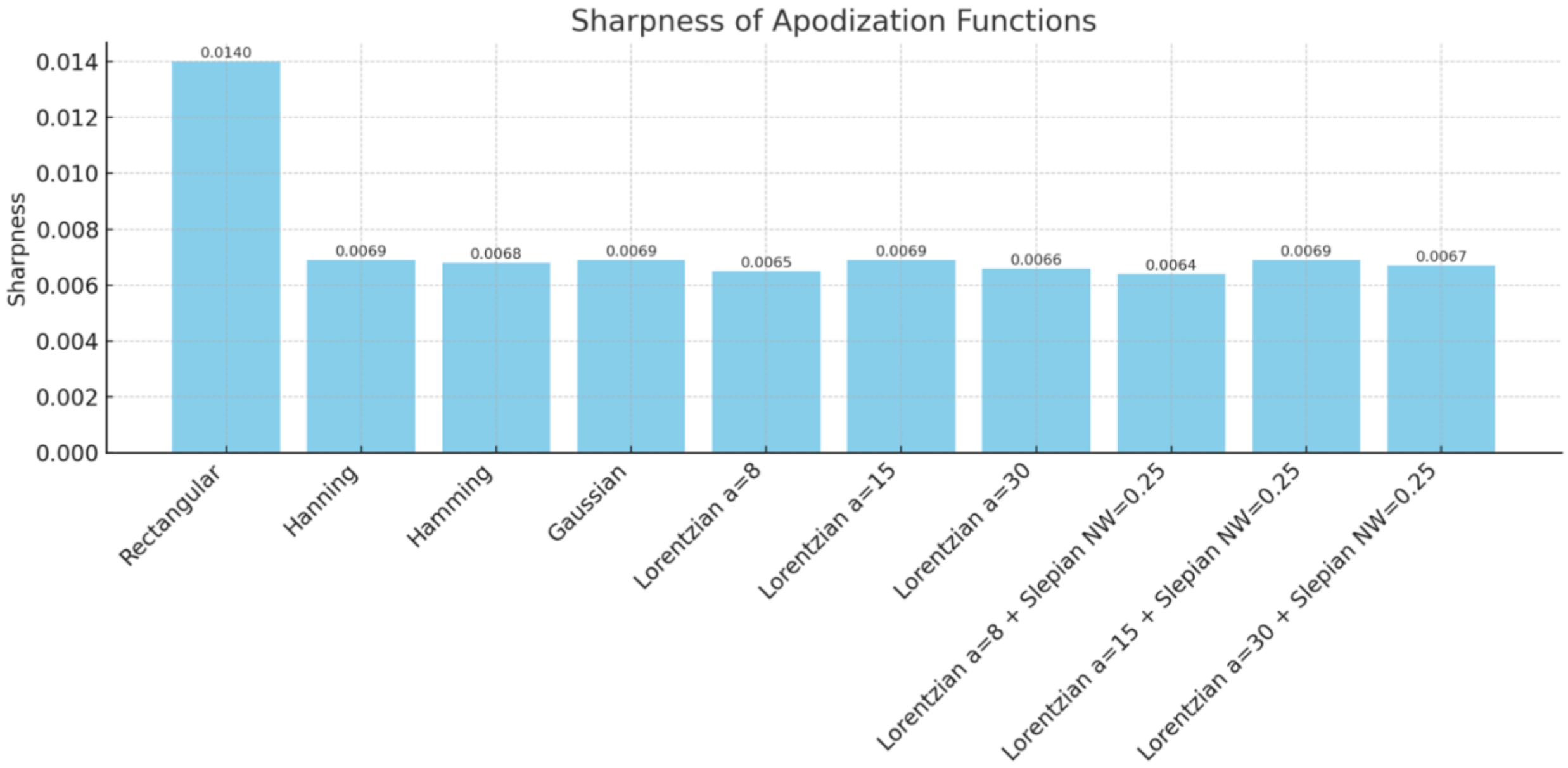

**Figure 8:** Illustrations shows comparisons of the A) Contrast to Noise ratio where the signal ROI (green ROI) is selected from the low contrast region and the background ROI (red ROI) and B) Sharpness of the contrast phantom for each of the different optimization functions used in Figure 7.

In this study, we evaluated the effects of different apodization functions on image quality metrics in ultrasound phantom imaging, focusing on contrast-to-noise ratio (CNR) and sharpness. Our analysis demonstrated that the choice of apodization substantially impacts both metrics. The rectangular function yielded the highest sharpness (0.014), reflecting strong edge preservation but at the expense of increased noise. In contrast, smoother windows such as Hanning, Hamming, and Gaussian produced lower sharpness values (~0.0068–0.0069), consistent with reduced sidelobe energy. CNR analysis of the low-contrast target revealed that certain Lorentzian-based windows, particularly Lorentzian a=8 and Lorentzian a=15, achieved the highest CNR values (>3.2), outperforming the original image and standard windows. Interestingly, the combination of Lorentzian functions with Slepian tapering generally reduced CNR, suggesting that excessive smoothing may compromise lesion detectability. These findings highlight the inherent trade-off between sharpness and CNR: while rectangular apodization optimizes sharpness, Lorentzian-based apodizations appear to provide the best balance by enhancing lesion conspicuity without excessively degrading spatial resolution.

## Discussion

Ultrasound imaging is a powerful diagnostic tool widely used in various medical specialties. However, like any imaging modality, ultrasound is susceptible to certain artifacts that can compromise image quality and interpretation. Reduced side lobes provided by this method would contribute to improvements in CNR in Ultrasound Imaging [7] Understanding these artifacts is crucial for clinicians and sonographers to make accurate diagnoses and avoid misinterpretations.

Grating lobes are artifacts that result from the use of phased array transducers [7]. Grating lobes occur when the ultrasound beam is steered at certain angles away from the main beam axis. Grating lobes are most prominent in large-aperture phased array transducers, and their presence is more noticeable in deep or far-field imaging.

To mitigate the effects of grating lobes, advanced beamforming algorithms, and signal processing techniques are employed. These methods aim to minimize the impact of secondary lobes, improve image clarity, and reduce false echoes.

Side lobes artifacts, also known as side lobes interference or side lobes clutter, are a type of artifact that arises from the transmission and reception of ultrasound waves [7]. They occur in all ultrasound imaging systems and result from the inherent nature of ultrasound beamforming.

Side lobes are the secondary lobes that appear adjacent to the main lobe of the ultrasound beam. These additional lobes are unavoidable due to the physics of wave propagation and the finite size of transducer elements. Side lobes can cause echoes from structures located away from the primary beam axis, leading to artifacts in the ultrasound image.

One common manifestation of side lobes artifacts is the appearance of ghost echoes or false reflections. These echoes can create duplicate structures in the image, making it challenging to accurately identify the true anatomical features. Side lobes artifacts are most pronounced in tissues with strong reflectors, such as bone interfaces, resulting in shadowing and masking of structures behind them.

To minimize side lobes artifacts, apodization techniques, as discussed in the previous abstract, are often employed. By applying windowing functions to the ultrasound signals, the amplitude of side lobes can be reduced, enhancing the main lobe resolution, and improving image quality. However, most apodization-based functions cause a widening of the main lobe which results in an increase in full width at half maximum reducing the lateral resolution in exchange for a reduction on side lobes. In this paper, we discuss a new apodization function that greatly reduces the sides lobes while reducing the full-width half maximum resulting in an increase in the lateral resolution of ultrasound. The Field II simulations show that the proposed Lorentzian apodization function achieve both maintenance of image sharpness and improve CNR. This is usually not possible to achieve with current apodization functions as reduction of side lobes improves the image CNR but often leads to a blurring of the image.

The sharper ultrasound image with a lower sidelobes will improve spatial resolution, lower artifacts, and increase tissue contrast, therefore, greatly benefit the clinical diagnosis accuracy. Spatial resolution, the ability to discriminate closely spaced structures, is essential for resolving fine anatomical features. Note that this helps, especially with small lesions detection differentiating between benign and malignant tumors, and some small vascular pathologies. Minimized unwanted echoes from graduated sidelobes that may mask or mimic actual anatomical features, lowering chance of false-positives or -negatives. Sidelobe artifacts, for example, can cause obscured borders of cystic or solid lesions and potentially misdiagnoses [10]. Better contrast resolution allows for clearer imaging of tissues that may share similar acoustic substations, and is valuable in visualizing subtle changes of pathology, including early hepatic fibrosis or microcalcifications of breast imaging.

Sharper imaging enables more precise measurements of tissue sizes, important for tracking such things as thyroid nodules or fetal development. It also improves visualization of small or deep structures, minimizing the scattering and sidelobes to ensure the imaging of deeper tissues such as the prostate or pancreas. The improved clarity provided by sharper ultrasound images leads to better operator independence, ensuring that diagnostic quality is maintained between sonographers. Moreover, enhanced beam profiles are particularly advantageous for advanced imaging paradigms such as elastography and contrast-enhanced ultrasound, which depend on accurate imaging to delineate tissue stiffness and vascular perfusion, respectively. In combination, these advances facilitate increased confidence and trust in ultrasound-based diagnostic procedures, which can ultimately result in more effective patient management and clinical outcome.

In routine ultrasound QA, beam profile quality, sidelobe suppression, and image contrast are essential parameters for verifying imaging performance and transducer integrity. This study demonstrates that the proposed Lorentzian apodization method significantly outperforms both rectangular and Hanning windows in terms of main beam FWHM, sidelobe suppression, wire phantom CNR, and sharpness ($p < 0.05$ for all metrics). These findings support the clinical potential of these apodization techniques for routine QA, where quantitative evaluation of beam performance and image quality is critical. The Lorentzian functions can be readily integrated into Field II simulations or programmable QA setups for evaluating lateral resolution and artifact suppression in ultrasound imaging systems.

## Conclusion

In conclusion, side lobes artifacts are common challenges in ultrasound imaging that arise due to the beamforming process. These artifacts can lead to misinterpretation and false diagnoses if not appropriately managed. As ultrasound technology continues to evolve, advanced beamforming algorithms, signal processing techniques, and apodization methods will play vital roles in minimizing these artifacts and improving the overall image quality. Additionally, continued research and innovation in transducer design and image reconstruction will contribute to further reducing these artifacts, ultimately enhancing the diagnostic accuracy and utility of ultrasound imaging in clinical practice.

**Acknowledgments**
The author(s) declare financial support was received for the research, authorship, and/or publication of this article.

The research in this publication was supported by the National GEM Consortium Fellowship and the SciMed GRS provided by the Graduate School, part of the Office of Vice Chancellor for Research and Graduate Education at the University of Wisconsin-Madison, with funding from the Wisconsin Alumni Research Foundation and the UW-Madison.

## Declaration of generative AI and AI-assisted technologies in the writing process

During the preparation of this work the author(s) used ChatGPT to provide concrete scientific literature that supported the results and for grammar corrections. After using this tool/service, the author(s) reviewed and edited the content as needed and take(s) full responsibility for the content of the publication.

## References

[1] Wagai T. Studies on the foundation and development of diagnostic ultrasound. Proc Jpn Acad Ser B Phys Biol Sci. 2007 Dec;83(8):256-65. doi: 10.2183/pjab/83.256. PMID: 24367150; PMCID: PMC3859294.
[2] Carovac A, Smajlovic F, Junuzovic D. Application of ultrasound in medicine. Acta Inform Med. 2011 Sep;19(3):168-71. doi: 10.5455/aim.2011.19.168-171. PMID: 23408755; PMCID: PMC3564184.
[3] Prabhu SJ, Kanal K, Bhargava P, Vaidya S, Dighe MK. Ultrasound artifacts: classification, applied physics with illustrations, and imaging appearances. Ultrasound Q. 2014 Jun;30(2):145-57. doi: 10.1097/RUQ.0b013e3182a80d34. PMID: 24850030.
[4] Seo CH, Yen JT. Sidelobe suppression in ultrasound imaging using dual apodization with cross-correlation. IEEE Trans Ultrason Ferroelectr Freq Control. 2008 Oct;55(10):2198-210. doi: 10.1109/TUFFC.919. PMID: 18986868; PMCID: PMC2905597.
[5] Ilovitsh, T., Ilovitsh, A., Foiret, J. *et al.* Acoustical structured illumination for super-resolution ultrasound imaging.*Commun Biol* **1**, 3 (2018). https://doi.org/10.1038/s42003-017-0003-5
[6] Guenther DA, Walker WF. Optimal apodization design for medical ultrasound using constrained least squares part I: theory. IEEE Trans Ultrason Ferroelectr Freq Control. 2007 Feb;54(2):332-42. doi: 10.1109/tuffc.2007.247. PMID: 17328330.
[7] Barthez PY, Léveillé R, Scrivani PV. Side lobes and grating lobes artifacts in ultrasound imaging. Vet Radiol Ultrasound. 1997 Sep-Oct;38(5):387-93. doi: 10.1111/j.1740-8261.1997.tb02104.x. PMID: 9335099.
[8] J.M. Thijssen, M. Mischi. X-ray and Ultrasound Imaging. *Comprehensive Biomedical Physics*, (2014)
[9] Thomas Szabo, Diagnostic Ultrasound Imaging. (2004)
[10] Gimber, L.H., Taljanovic, M.S. (2017). Ultrasound Imaging Artifacts. In: Peh, W. (eds) Pitfalls in Musculoskeletal Radiology. Springer, Cham. https://doi.org/10.1007/978-3-319-53496-1_2